\documentclass[10pt,conference]{IEEEtran}
\IEEEoverridecommandlockouts
\usepackage{cite}
\usepackage{amsmath,amssymb,amsfonts}
\usepackage{mathtools}
\usepackage{algorithmic}
\usepackage{graphicx}
\usepackage{textcomp}
\usepackage{xcolor}
\usepackage{verbatim}
\usepackage{physics}
\usepackage{subcaption}

\def\ddefloop#1{\ifx\ddefloop#1\else\ddef{#1}\expandafter\ddefloop\fi}

\def\ddef#1{\expandafter\def\csname b#1\endcsname{\ensuremath{\boldsymbol{#1}}}}
\ddefloop ABCDEFGHIJKLMNOPQRSTUVWXYZabcdefghijklmnopqrstuvwxyz\ddefloop

\def\ddef#1{\expandafter\def\csname c#1\endcsname{\ensuremath{\mathcal{#1}}}}
\ddefloop ABCDEFGHIJKLMNOPQRSTUVWXYZ\ddefloop

\def\ddef#1{\expandafter\def\csname s#1\endcsname{\ensuremath{\mathsf{#1}}}}
\ddefloop ABCDEFGHIJKLMNOPQRSTUVWXYZ\ddefloop

\def\Reals{{\mathbb R}} 
\def\Complex{{\mathbb C}} 
\def\dif{{\mathrm{d}}} 
\def\p{{\partial}} 
\def\ad{\text{ad}} 
\def\Ad{\text{Ad}} 
\def\su{\mathfrak{su}} 
\def\SU{\text{SU}} 
\DeclareMathOperator*{\argmax}{\text{argmax}}

\newtheorem{theorem}{Theorem}

\begin{document}

\title{Constructing Noise-Robust Quantum Gates via Pontryagin's Maximum Principle\\
\thanks{This work was supported by the Department of Energy under grant no. DE-SC0022389.}}

\author{\IEEEauthorblockN{Joshua Hanson}
\IEEEauthorblockA{\textit{Error Corp.} \\
Urbana IL, USA \\
josh@error-corp.com}
\and
\IEEEauthorblockN{Dennis Lucarelli}
\IEEEauthorblockA{\textit{Error Corp.} \\
College Park MD, USA \\
dennis@error-corp.com}
}

\maketitle

\begin{abstract}

Reliable quantum information technologies depend on precise actuation and techniques to mitigate the effects of undesired disturbances such as environmental noise and imperfect calibration. In this work, we present a general framework based in geometric optimal control theory to synthesize smooth control pulses for implementing arbitrary noise-robust quantum gates. The methodology applies to generic unitary quantum dynamics with any number of qubits or energy levels, any number of control fields, and any number of disturbances, extending existing dynamical decoupling approaches that are only applicable for limited gate sets or small systems affected by one or two disturbances. The noise-suppressing controls are computed via indirect trajectory optimization based on Pontryagin's maximum principle, eliminating the need to make heuristic structural assumptions on parameterized pulse envelopes.

\end{abstract}


\section{Introduction}\label{sec:intro}

Quantum devices used in near-term quantum computing are subject to noise processes that degrade the fidelity of the quantum gates used to execute a quantum computational task. Errors due to imperfect quantum gates may accumulate during the course of a computation and corrupt the measurement results containing the output of the quantum algorithm. This loss of quantum coherence severely limits both the number of computational qubits and the depth of the quantum circuit implementing the algorithm. 

Several complementary scientific and engineering methodologies have been developed to enhance coherence and mitigate noise in quantum devices, including improvements in materials, device design, and post-processing error mitigation. In addition to these, advanced control techniques may be employed to decouple the quantum state from its environment.  Control protocols designed to isolate a quantum system from its environment trace their origin to refocusing techniques in nuclear magnetic resonance and have been extensively studied within the context of quantum information processing. Dynamical decoupling, originally proposed for protecting idling operations \cite{viola1999dynamical} and subsequently generalized to construct {\it dynamically corrected} single-qubit gates \cite{khodjasteh2009dynamically},  provide families of gate sequences for noise-suppression that may be implemented at the circuit level. These approaches, however, often require strong and fast impulse controls which may violate bandwidth and/or amplitude constraints imposed by the electronics hardware implementing the control. Thus, developing noise-suppression methods based on smooth, physically realizable control pulses is an active area of research. Space Curve Quantum Control (SCQC) is one such framework that aims to explicitly construct closed ``error curves'' that quantify the defect between noise-affected and ideal noiseless evolution operators \cite{zeng2019geometric, buterakos2021geometrical}. Smooth, noise-suppressing controls can be extracted based on the extrinsic geometry (e.g., curvature and torsion) of these error curves. The SCQC formalism provides inspiration for this work, which is based on a similar perturbative analysis via the Magnus expansion but utilizes different techniques to synthesize the control.

In this paper we address the quantum noise-suppression control problem using the machinery of nonlinear optimal control and Pontryagin's Maximum Principle (PMP). PMP provides first-order necessary conditions for the minimization of a cost functional subject to a dynamical state equation. For general nonlinear systems, these conditions result in a nonlinear differential equation for the so-called {\it co-state}, which may be difficult to solve.  The popular quantum control technique known as GRAPE \cite{khaneja2005optimal} can be viewed in the framework of PMP and circumvents this problem by removing any constraints on the control in the cost function, effectively obtaining a constant co-state equation.  In contrast, our approach obtains tractable conditions for optimality which can be solved using nonlinear constrained optimization. The use of PMP for constructing smooth time-optimal and robust control pulses for Rydberg atoms is also explored in \cite{pupillo2022timeoptimal}.

The paper is organized as follows: In Section \ref{sec:problem} we formulate the control problem for noise-robust gate design and state the PMP necessary conditions. In Section \ref{sec:examples} we specialize the control problem for two single-qubit examples, one of which incorporates two independent noise sources and the other demonstrates higher order disturbance rejection. Results from numerical simulation for these two examples are shown in Section \ref{sec:results}, with conclusions and discussion of follow-up work provided in Section \ref{sec:discussion}.

\section{Formulating the control problem}\label{sec:problem}

Consider an $N$-dimensional quantum system described by a separable time-dependent Hamiltonian
\begin{equation}\label{eq:total_hamiltonian}
    H(t) = H^{(0)}(t) + \epsilon_1 H^{(1)}(t) + \cdots + \epsilon_n H^{(n)}(t),
\end{equation}
where $\epsilon_1,\dots,\epsilon_n \in \Reals$ are unknown constants representing various disturbances or quasi-static noise sources. We interpret $H^{(0)}$ as the ideal control Hamiltonian and $H^{(1)},\dots,H^{(n)}$ as noise generators or other perturbations whose contribution to the total Hamiltonian $H$ is unknown and could vary over multiple experiments. Suppose each term $H^{(i)}$ comprises a drift component plus a component that is linear in $m$ control inputs $u_1,\dots,u_m : [0,T] \to \Reals$, which we will express as
\begin{equation}\label{eq:hamiltonian_terms}
    H^{(i)}(t) = H_0^{(i)} + \sum_{j=1}^m u_j(t) H_j^{(i)},\quad 0 \leq i \leq n.
\end{equation}

Without loss of generality, we will assume $H$ has zero trace. The time-dependent unitary evolution operator $U \in \SU(N)$ for this system is governed by the Schr\"odinger equation
\begin{equation}\label{eq:schrodinger}
    \dot{U} = {-i} H U,\quad\quad U(0) = I,
\end{equation}
where the controls $u_1,\dots,u_m$ and disturbances $\epsilon_1,\dots,\epsilon_n$ enter through $H$ as in \eqref{eq:total_hamiltonian} and \eqref{eq:hamiltonian_terms}. We say a control pulse implements a particular unitary operation or \textit{gate} $G \in \SU(N)$ if $U(T) = G$ whenever $\epsilon_1 = \cdots = \epsilon_n = 0$. The noise-robust gate design problem is to construct control signals $u_1,\dots,u_m$ to implement a gate $G$ in such a way that the gate fidelity
\begin{equation}\label{eq:fidelity}
    \cF = \frac{1}{N} \left| \text{tr}\left( G^\dagger \cdot U(T) \right) \right|,
\end{equation}
which measures the phase-invariant distance between the target gate $G$ and the evolution operator $U(T)$ at the terminal time, is locally invariant up to some order with respect to the disturbances $\epsilon_1,\dots,\epsilon_n$. We can quantify robustness as follows: We say a control pulse is \textit{robust} to order $r$ if, in the limit of vanishing disturbances, the ideal fidelity is unity and the partial derivatives up to order $r$ of the fidelity with respect to the disturbances vanish. Mathematically, this means
\begin{align}
    \cF|_{\epsilon_1 = \cdots = \epsilon_n = 0} &= 1 \label{eq:fidelity_robustness} \\
    \frac{\p^k \cF}{\p \epsilon_{i_1} \cdots \p \epsilon_{i_k}}\bigg|_{\epsilon_1 = \cdots = \epsilon_n = 0} &= 0 \label{eq:fidelity_derivatives_robustness}
\end{align}
for $1 \leq k \leq r$ and $1 \leq i_1,\dots,i_k \leq n$. Using an appropriate perturbative expansion of the evolution operator, we can phrase this robust control problem as an equivalent point-to-point motion planning problem, which can be solved using techniques from geometric optimal control theory. In the following subsections we will construct the augmented control system and state this motion planning problem, then describe how it can be upgraded to an optimal control problem and solved using Pontryagin's maximum principle.

\subsection{The augmented control system}

Treating the terms $\epsilon_1 H^{(1)}, \dots, \epsilon_n H^{(n)}$ as perturbations to the ideal control Hamiltonian $H^{(0)}$, consider the evolution operator in the interaction picture. Substituting the decomposition $U = RU_I$ into \eqref{eq:schrodinger}, we obtain the governing equations for the ideal evolution operator $R$ and the evolution operator in the interaction picture $U_I$, respectively, as
\begin{alignat}{3}
    \dot{R} &= {-i} H^{(0)} R,&&\quad\quad R(0) &&= I, \label{eq:R_dynamics} \\
    \dot{U}_I &= {-i} H_I U_I,&&\quad\quad U_I(0) &&= I, \label{eq:U_I_dynamics}
\end{alignat}
where the interaction Hamiltonian $H_I$ is given by
\begin{equation*}
    H_I(t) = \Ad_{R^\dagger} \left( {-i} \sum_{i=1}^n \epsilon_i H^{(i)}(t) \right)
\end{equation*}
and $\Ad : \SU(N) \to \text{Aut}(\su(N))$, $\Ad_{R^\dagger}(H) = R^\dagger H R$ is the adjoint action of $\SU(N)$. Writing $U_I = \exp(\Omega)$, we can pullback the dynamics \eqref{eq:U_I_dynamics} for $U_I$ on $\SU(N)$ to dynamics for $\Omega$ on $\su(N)$, yielding
\begin{equation*}
    \dot{\Omega} = \dif\exp^{-1}_\Omega \left( {-i H_I} \right) = \sum_{k=0}^\infty \frac{B_k}{k!} \ad^k_\Omega \left( {-i H_I} \right),\quad \Omega(0) = 0,
\end{equation*}
where $\{B_k\}_{k=0}^\infty$ are the Bernoulli numbers with $B_1 = {-\frac{1}{2}}$ and $\ad : \su(N) \to \text{Der}(\su(N))$, $\ad_\Omega = \dif(\Ad)_I(\Omega)$ is the adjoint action of $\su(N)$, where $I \in \SU(N)$ is the identity. The iterated adjoint operator $\ad^k_\Omega : \su(N) \to \su(N)$ satisfies the recursive formula
\begin{align*}
    \ad^0_\Omega H &= H \\
    \ad^{k}_\Omega H &= [\Omega, \ad^{k-1}_\Omega H]
\end{align*}
where $[X,Y] = XY - YX$ is the standard Lie bracket on $\su(N)$. We call $\Omega \in \su(N)$ an \textit{error curve} because it can be visualized as a space curve in the Lie algebra $\su(N)$ and its magnitude quantifies the logarithmic distance between the ideal evolution operator $R$ and the true evolution operator $U$ \cite{barnes2022dynamically}. Taking advantage of the fact that $\su(N)$ is a vector space, we can represent $\Omega$ as an infinite series
\begin{equation*}
    \Omega(t) = \sum_{k=1}^\infty \Omega_k(t)
\end{equation*}
and define dynamics for each term individually via the initialization $\dot{\Omega}_1 = {-i H_I}$ and the recursive formula
\begin{equation*}
    \dot{\Omega}_k = \sum_{\ell=1}^{k-1} \frac{B_\ell}{\ell!} \sum_{k_1 + \cdots + k_\ell = k-1} \ad_{\Omega_{k_1}} \cdots \ad_{\Omega_{k_\ell}} \left( {-i H_I} \right)
\end{equation*}
for $k \geq 2$. The first few terms satisfy
\begin{align*}
    \dot{\Omega}_1 &= {-i H_I} \\
    \dot{\Omega}_2 &= {-\frac{1}{2}} \big[\Omega_1, {-i H_I} \big] \\
    \dot{\Omega}_3 &= {-\frac{1}{2}} \big[\Omega_2, {-i H_I} \big] + \frac{1}{12} \big[\Omega_1, \big[\Omega_1, {-i H_I} \big]\big].
\end{align*}
Integrating each term yields the so-called \textit{Magnus series} \cite{magnus1954exponential}, however retaining the presentation as matrix ODEs will be critical to formulating the motion planning problem later on. Further taking advantage of the fact that $H_I$ is linear in the disturbances $\epsilon_1,\dots,\epsilon_n$, we can decompose each term $\Omega_k$ as
\begin{equation*}
    \Omega_k(t) = \sum_{i_1,\dots,i_k=1}^n \epsilon_{i_1} \cdots \epsilon_{i_k} \Omega_k^{(i_1,\dots,i_k)}(t).
\end{equation*}
We can express the dynamics for each new term using a similar recursive procedure, yielding $\dot{\Omega}_1^{(i_1)} = \Ad_{R^\dagger} \left( {-i H^{(i_1)}} \right)$ and
\begin{align}
    &\dot{\Omega}_k^{(i_1,\dots,i_k)} = \sum_{\ell=1}^{k-1} \frac{B_\ell}{\ell!} \sum_{k_1 + \cdots + k_\ell = k-1} \label{eq:Omega_dynamics} \\
    &\quad\quad \ad_{\Omega_{k_1}^{(i_1,\dots,i_{k_1})}} \cdots \ad_{\Omega_{k_\ell}^{(i_{k-k_\ell},\dots,i_{k-1})}} \Ad_{R^\dagger} \left( {-i H^{(i_k)}} \right) \nonumber
\end{align}
for $k \geq 2$ and $1 \leq i_1,\dots,i_k \leq n$. The first few terms satisfy
\begin{align*}
    \dot{\Omega}^{(i_1)}_1 &= {-i R^\dagger H^{(i_1)} R} \\
    \dot{\Omega}^{(i_1,i_2)}_2 &= {-\frac{1}{2}} \big[\Omega^{(i_1)}_1, {-i R^\dagger H^{(i_2)} R} \big] \\
    \dot{\Omega}^{(i_1,i_2,i_3)}_3 &= {-\frac{1}{2}} \big[\Omega^{(i_1,i_2)}_2, {-i R^\dagger H^{(i_3)} R} \big] \\
    &\quad\quad + \frac{1}{12} \big[\Omega^{(i_1)}_1, \big[\Omega^{(i_2)}_1, {-i R^\dagger H^{(i_3)} R} \big]\big].
\end{align*}
Notice that the differential equations above no longer depend on the disturbances $\epsilon_1,\dots,\epsilon_n$, which we will exploit to construct robust control pulses. For $n$ disturbances and robustness order $r$, there are $p = n + n^2 + \cdots + n^r$ independent error curves $\Omega_{k}^{(i_1,\dots,i_k)}$, each of which describes the contribution of the monomial $\epsilon_{i_1} \cdots \epsilon_{i_k}$ to the total error curve $\Omega$.

\subsection{The motion planning problem}

Recall the definition of robustness stated earlier in this section. The condition \eqref{eq:fidelity_robustness} is equivalent (up to a global phase) to the final state constraint $R(T) = G$, and the insensitivity conditions \eqref{eq:fidelity_derivatives_robustness} will be satisfied if $\Omega_{k}^{(i_1,\dots,i_k)}(T) = 0$ for $1 \leq k \leq r$ and $1 \leq i_1,\dots,i_k \leq n$, which can be verified by substituting $U(T) = R(T) \exp(\Omega(T))$ into \eqref{eq:fidelity} and evaluating \eqref{eq:fidelity_robustness} and \eqref{eq:fidelity_derivatives_robustness}. This allows us to phrase the noise-robust gate design problem as a point-to-point motion planning problem as follows: We seek an open-loop control policy $u : [0,T] \to \Reals^m$ to steer the system governed by \eqref{eq:R_dynamics} and \eqref{eq:Omega_dynamics}, which evolves on the product manifold $\cM = \SU(N) \times \su(N)^{p}$, from the initial configuration
\begin{equation}\label{eq:initial_constraint}
    R(0) = I,\quad\quad \Omega_{k}^{(i_1,\dots,i_k)}(0) = 0
\end{equation}
at time 0 to the final configuration
\begin{equation}\label{eq:final_constraint}
    R(T) = G,\quad\quad \Omega_{k}^{(i_1,\dots,i_k)}(T) = 0
\end{equation}
at time $T$, for $1 \leq k \leq r$ and $1 \leq i_1,\dots,i_k \leq n$. Some care should be taken to ensure that the augmented system is indeed globally controllable, which may introduce practical limits on $n$ and $r$. Studying the reachability of this system for different control Hamiltonians and noise generators is an interesting direction for follow-up study. For the remainder of this work we will assume the control system \eqref{eq:R_dynamics} and \eqref{eq:Omega_dynamics} is globally controllable.

In general there are several, if not infinitely many, possible control pulses that will satisfy the above dynamical and endpoint constraints. To discriminate between them, we upgrade the motion planning problem above to a free-time fixed-endpoint optimal control problem. Given a running cost $\cL : \cM \times \Reals^m \to \Reals$, we seek a control policy $u : [0,T] \to \Reals^m$ that minimizes the cost functional
\begin{equation}\label{eq:cost_functional}
    J(u) = \int_0^T \cL(x(t),u(t)) \dif t,
\end{equation}
where the state trajectory $x : [0,T] \to \cM$ is subject to the dynamics \eqref{eq:R_dynamics} and \eqref{eq:Omega_dynamics} and the initial and final endpoint constraints \eqref{eq:initial_constraint} and \eqref{eq:final_constraint}, respectively. Generally the terminal time $T$ must be free, unless the drift components in the control and perturbation Hamiltonians satisfy $H_0^{(i)} = 0$ for $0 \leq i \leq n$. We can address this by introducing an additional fictitious constant control $u_0 \in \Reals$ multiplying each $H_0^{(i)}$ for $0 \leq i \leq n$ and setting $T = 1$. Then we can simply re-scale the controls $u_0,u_1,\dots,u_m$ as necessary during a subsequent calibration procedure given the desired time-horizon and actual drift coefficient.

\subsection{Pontryagin's maximum principle}\label{ssec:pmp}

We must transcribe this infinite-dimensional optimization problem into a finite-dimensional optimization problem so that it can be solved numerically. Many previous works \cite{SKINNER2010248,lucarelli2018quantum,oda2023optimally,sorensen2018quantum,machnes2018tunable} use direct transcription methods where the controls $u_1,\dots,u_m$ are parameterized as linear combinations of some temporal basis functions $\{v_k : [0,T] \to \Reals\}_{k=1}^M$
\begin{equation*}
    u_j(t) = \sum_{k=1}^M \alpha_{j}^k v_k(t).
\end{equation*}
The transcribed objective is to find the coefficients $\alpha_j^k \in \Reals$, $1 \leq j \leq m$, $1 \leq k \leq M$ that minimize $J$ subject to the dynamical and endpoint constraints.

In this work, we instead apply an indirect transcription method which states necessary conditions for optimality prior to discretization. We then discretize the optimality conditions directly and solve them numerically to obtain a candidate optimal control policy. Pontryagin's maximum principle furnishes such necessary conditions --- we state the applicable version of the maximum principle below (cf. \cite{chang2011simple})

\begin{theorem}[Pontryagin's maximum principle on manifolds]

Consider a control system $\dot{x} = f(x,u)$, $x \in \cM$ subject to initial and final endpoint constraints $x(0) \in \cS_0$, $x(T) \in \cS_1$. Suppose $u^* : [0,T] \to \Reals^m$ is an optimal control policy in the sense of minimizing \eqref{eq:cost_functional} and $x^* : [0,T] \to \cM$ is the resulting optimal state trajectory. Then there exists a costate trajectory $p^*(t) \in T^*_{x^*(t)} \cM$ and a constant $p_0^* \leq 0$ satisfying $(p_0^*, p^*(t)) \neq (0,0)$ for all $t \in [0,T]$ such that the following conditions hold:
\begin{itemize}
    \item The state and costate trajectory $(x^*,p^*) : [0,T] \to T^* \cM$ is an integral curve of the Hamiltonian vector field ${X_\cH : T^* \cM \to T(T^* \cM)}$ defined by
    \begin{equation}\label{eq:pmp_hamiltonian_vector_field}
        \dif \cH(Y) = \omega(X_\cH, Y)
    \end{equation}
    for every vector field $Y : T^* \cM \to T(T^* \cM)$. Here $\omega : T(T^* \cM) \times T(T^* \cM) \to \Reals$ is the canonical symplectic form on the cotangent bundle $T^* \cM$. The control Hamiltonian $\cH : T^* \cM \to \Reals$, not to be confused with the Hamiltonian \eqref{eq:total_hamiltonian}, is defined as
    \begin{equation}\label{eq:pmp_hamiltonian}
        \cH(x,p; p_0,u) = p(f(x,u)) + p_0 \cL(x,u).
    \end{equation}
    \item For each $t \in [0,T]$, the value of the optimal control $u^*(t)$ maximizes the Hamiltonian:
    \begin{equation*}
        \cH(x^*(t),p^*(t); p_0^*,u^*(t)) = \max_{u \in \Reals^m} \cH(x^*(t),p^*(t); p_0^*,u).
    \end{equation*}
    \item The Hamiltonian vanishes along the optimal trajectory:
    \begin{equation*}
        \cH(x^*(t),p^*(t); p_0^*,u^*(t)) \equiv 0.
    \end{equation*}
    \item The initial costate is orthogonal to the tangent space to $\cS_0$ at $x^*(0)$ and the final costate is orthogonal to the tangent space to $\cS_1$ at $x^*(T)$:
    \begin{equation*}
        p^*(0)(\gamma_0) = 0,\quad p^*(T)(\gamma_1) = 0
    \end{equation*}
    for every $\gamma_0 \in T_{x^*(0)} \cS_0$ and every $\gamma_1 \in T_{x^*(T)} \cS_1$.
\end{itemize}

\end{theorem}

Generally we normalize the costate so that $p_0^* \equiv -1$, which is always possible provided that $p_0^* \neq 0$, and we assume that this holds for the remainder of this work (i.e., we will ignore singular controls). Since we are working with a fixed-endpoint problem, the initial and final endpoint constraint sets are given by $\cS_0 = \{x_0\}$ and $\cS_1 = \{x_1\}$, where the points $x_0$ and $x_1$ are specified by \eqref{eq:initial_constraint} and \eqref{eq:final_constraint}. This implies that the tangent spaces $T_{x^*(0)} \cS_0$ and $T_{x^*(T)} \cS_1$ in the transversality conditions are empty, and thus there are no constraints imposed on the initial and final costates $p^*(0)$ and $p^*(T)$.

Since the costate is finite-dimensional, we can transcribe the optimization problem as follows: We seek an initial costate $p^*(0) \in T^*_{x^*(0)} \cM$ such that the candidate optimal control policy $u^* : [0,T] \to \Reals^m$ --- computed pointwise along the state and costate trajectory $(x^*,p^*)$ by maximizing the control Hamiltonian $\cH$ defined in \eqref{eq:pmp_hamiltonian} --- steers the initial state $x^*(0) = x_0$ to the final state $x^*(T) = x_1$. This transcription is the so-called \textit{single shooting method}. Modifying this transcription to a \textit{multiple shooting method} by introducing additional degrees of freedom along the state and costate trajectory and subsequently removing them by adding a matching number of continuity constraints often significantly improves numerical stability of the optimization problem. Alternatively, collocation methods based on parameterizing the candidate state and costate trajectories as a sum of basis functions and imposing the dynamical constraints at a carefully chosen set of collocation points throughout the time horizon is another strategy for improving regularity of the optimization problem. For any transcription the derivation of the state and costate equations remains the same, so the transcription method and numerical optimization algorithm can be chosen based on which yields the fastest and/or most consistent convergence.

To solve the finite-dimensional optimization problem resulting from the single shooting transcription numerically, we define an objective function that computes the magnitude of the residual between the desired final configuration \eqref{eq:final_constraint} and the one that results from the candidate control policy, which is computed by integrating the Hamiltonian vector field $X_\cH$ defined in \eqref{eq:pmp_hamiltonian_vector_field} numerically. We can apply standard nonlinear programming algorithms to minimize this objective function, which will yield feasible candidate optimal control policies. We do not need to explicitly minimize \eqref{eq:cost_functional} in this program because this is already built into the necessary conditions from PMP. However in the absence of sufficient conditions for optimality, we may need to compare different candidate controls proposed by some globalization method, such as seeding random initial guesses to the numerical algorithm and selecting the feasible control resulting in the smallest cost.

\subsection{Right-trivialized Hamiltonian dynamics}

In order to solve the transcribed optimization problem stated above, we wish to write down a system of matrix differential equations that represents the Hamiltonian vector field $X_\cH$. To reflect the natural geometry of the problem, we will compute a representation of the right-trivialization of $X_\cH$, which simplifies the costate dynamics. We can equip the Lie algebra $\mathfrak{su}(N)$ --- represented as the set of traceless skew-Hermitian matrices --- with a natural inner product given by $\langle X, Y \big\rangle = \frac{1}{N} \text{tr}(X^\dagger Y)$. This inner product can be used to canonically identify $\mathfrak{su}(N)$ with its dual space $\mathfrak{su}(N)^*$ via $\mu \leftrightarrow \big\langle \mu, \cdot \big\rangle$, which we will apply in the sequel.

Here we review some necessary constructions from differential geometry. The cotangent bundle $T^* G$ of a Lie group $G$ with Lie algebra $\mathfrak{g}$ is isomorphic to the trivial cotangent bundle $G \times \mathfrak{g}^*$. We can construct a bundle isomorphism using the right-multiplication map. Define
\begin{equation*}
    \phi : G \times \mathfrak{g}^* \to T^* G,\quad (g,\mu) \mapsto \left( g, \left( \dif(R_{g^{-1}})_g \right)^* \mu \right),
\end{equation*}
where
\begin{equation*}
    \dif(R_{g^{-1}})_g : T_g G \to \mathfrak{g},\quad X \mapsto Xg^{-1}
\end{equation*}
is the differential at $g \in G$ of the right-multiplication map
\begin{equation*}
    R_{g^{-1}} : G \to G,\quad h \mapsto hg^{-1}
\end{equation*}
and $\left( \dif(R_{g^{-1}})_g \right)^* : \mathfrak{g}^* \to T_g^* G$ is its algebraic adjoint. The canonical symplectic form $\omega : T(T^* G) \times T(T^* G) \to \Reals$ on the cotangent bundle $T^* G$ is given by $\omega = -\dif \theta$, where
\begin{equation*}
    \theta : T^* G \to T^*(T^* G),\quad \theta_{(g,p)} = p \circ \dif \pi_{(g,p)}
\end{equation*}
is the tautological one-form and $\pi : T^* G \to G$, $\pi(g,p) = g$ is the usual cotangent bundle projection. Pulling back $\theta$ by $\phi$ and evaluating at $(X,\alpha) \in T_{(g,\mu)}(G \times \mathfrak{g}^*)$ gives the expression
\begin{equation*}
    \phi^* \theta_{(g,\mu)}(X,\alpha) = \mu(X).
\end{equation*}
Taking the negative differential of $\phi^* \theta$ gives the right-trivialized symplectic form
\begin{equation*}
    \phi^* \omega_{(g, \mu)}( (\Xi, \alpha), (\Upsilon, \beta) ) = \beta(\Xi) - \alpha(\Upsilon) - \mu([\Xi, \Upsilon]),
\end{equation*}
which follows from some straightforward calculations. The cotangent bundle $T^* \mathfrak{g}$ of a Lie algebra $\mathfrak{g}$ is naturally isomorphic to the product space $\mathfrak{g} \times \mathfrak{g}^*$ --- denote this isomorphism by $\psi : \mathfrak{g} \times \mathfrak{g}^* \to T^* \mathfrak{g}$. The pullback by $\psi$ of the canonical symplectic form on $T^* \mathfrak{g}$ is simply
\begin{equation*}
    \psi^* \omega( (\Xi, \alpha), (\Upsilon, \beta) ) = \beta(\Xi) - \alpha(\Upsilon),
\end{equation*}
which holds at all points in $\mathfrak{g} \times \mathfrak{g}^*$. Lastly, the canonical symplectic form on the cotangent bundle of a product manifold $\cM_1 \times \cM_2$ is equal to $\pi_1^* \omega_1 + \pi_2^* \omega_2$, where $\omega_1$ and $\omega_2$ are the canonical symplectic forms on $T^* \cM_1$ and $T^* \cM_2$, and $\pi_i : \cM_1 \times \cM_2 \to \cM_i$ is the projection onto $\cM_i$ for $i=1,2$.

Now let $G = \SU(N)$ and $\mathfrak{g} = \su(N)$. Pulling back the control Hamiltonian
\begin{equation*}
    \cH : T^* \left( \SU(N) \times \mathfrak{su}(N)^{p} \right) \to \Reals
\end{equation*}
by $\phi \times \psi^{p}$ gives the right-trivialized control Hamiltonian
\begin{gather*}
    \widehat{\cH} : \SU(N) \times \su(N)^{2p+1} \to \Reals \\
    (R, \dots, \Omega_k^{(i_1,\dots,i_k)}, \dots, \mu_R, \dots, \mu_k^{(i_1,\dots,i_k)}, \dots) \mapsto \quad\quad\quad\quad \\
    \big\langle \mu_R, {-i H^{(0)}} \big\rangle + \sum_{k=1}^r \sum_{i_1,\dots,i_k = 1}^n \big\langle \mu_k^{(i_1,\dots,i_k)}, \dot{\Omega}_k^{(i_1,\dots,i_k)} \big\rangle - \cL,
\end{gather*}
where we have identified each copy of $\su(N)^*$ with $\su(N)$. We can compute the right-trivialized Hamiltonian vector field
\begin{equation*}
    X_{\widehat{\cH}} : \SU(N) \times \mathfrak{su}(N)^{2p+1} \to T \left( \SU(N) \times \mathfrak{su}(N)^{2p+1} \right)
\end{equation*}
directly from the definition in the maximum principle
\begin{equation*}
    \dif \widehat{\cH}(\cdot) = \widehat{\omega}(X_{\widehat{\cH}}, \cdot),
\end{equation*}
where we denote the right-trivialized symplectic form by $\widehat{\omega}$. We derive a system of matrix differential equations governing the integral curves by evaluating the vector field $X_{\widehat{\cH}}$ on a test trajectory in $\SU(N) \times \mathfrak{su}(N)^{2p+1}$. Half of the equations recover the state dynamics \eqref{eq:R_dynamics} and \eqref{eq:Omega_dynamics}. While it is possible to write generic closed-form expressions for the costate dynamics as well, they are very cumbersome, so we will instead demonstrate the calculations through some illustrative examples in the following section.

\section{Examples}\label{sec:examples}

We use the following notation for the Pauli matrices
\begin{equation*}
    X = \begin{bmatrix} 0 & 1 \\ 1 & 0 \end{bmatrix},\quad Y = \begin{bmatrix} 0 & {-i} \\ i & 0 \end{bmatrix},\quad Z = \begin{bmatrix} 1 & 0 \\ 0 & {-1} \end{bmatrix}.
\end{equation*}
We will also suppress the notation $\widehat{\cdot}$ related to the right-trivialization of the Hamiltonian and symplectic form.

\subsection*{Example 1}

Consider a single-qubit system ($N=2$) driven at its resonance frequency with in-phase and quadrature controls, subject to dephasing and control amplitude scaling errors. The total Hamiltonian in the rotating-wave frame is given by
\begin{equation*}
    H(t) = (1 + \epsilon_2) \left( u_1(t) X + u_2(t) Y \right) + \epsilon_1 Z.
\end{equation*}
We have $n = 2$ disturbances and $m = 2$ controls, and the components of the Hamiltonian are given by
\begin{alignat*}{5}
    H_0^{(0)} &= 0,\quad &&H_1^{(0)} &&= X,\quad &&H_2^{(0)} &&= Y, \\
    H_0^{(1)} &= Z,\quad &&H_1^{(1)} &&= 0,\quad &&H_2^{(1)} &&= 0, \\
    H_0^{(2)} &= 0,\quad &&H_1^{(2)} &&= X,\quad &&H_2^{(2)} &&= Y.
\end{alignat*}
We seek controls $u_1,u_2$ that are robust to first order ($r = 1$)
to implement a Hadamard gate
\begin{equation*}
    G = \frac{i}{\sqrt{2}} \begin{bmatrix} 1 & 1 \\ 1 & -1 \end{bmatrix}
\end{equation*}
while minimizing the cost functional
\begin{equation*}
    J = \frac{1}{2} \int_0^T u_1(t)^2 + u_2(t)^2 \dif t.
\end{equation*}
The control Hamiltonian is given by
\begin{align*}
    \cH &= \big\langle \mu_R, {-i \left( u_1 X + u_2 Y \right)} \big\rangle + \big\langle \mu_1^{(1)}, {-i R^\dagger Z R} \big\rangle \\
    &\quad + \big\langle \mu_1^{(2)}, {-i R^\dagger \left( u_1 X + u_2 Y \right) R} \big\rangle - \frac{1}{2} u_1^2 - \frac{1}{2} u_2^2.
\end{align*}
To compute the state and costate dynamics, we will first evaluate the differential of $\cH$ at the point
\begin{equation*}
    (x,\mu) = \Big( R,\Omega_1^{(1)},\Omega_1^{(2)},\mu_R,\mu_1^{(1)},\mu_1^{(2)} \Big) \in T^* \cM
\end{equation*}
on an arbitrary tangent vector
\begin{equation*}
    \eta = \Big(\Upsilon_R, \Upsilon_1^{(1)}, \Upsilon_1^{(2)}, \beta_R,\beta_1^{(1)},\beta_1^{(2)}\Big) \in T_{(x,\mu)}(T^* \cM).
\end{equation*}
This gives
\begin{align*}
    \dif \cH_{(x,\mu)}(\eta) &= \big\langle \beta_R, {-i \left( u_1 X + u_2 Y \right)} \big\rangle + \big\langle \beta_1^{(1)}, {-i R^\dagger Z R} \big\rangle \\
    &\quad + \big\langle \beta_1^{(2)}, {-i R^\dagger \left( u_1 X + u_2 Y \right) R} \big\rangle \\
    &\quad + \big\langle \mu_1^{(1)}, R^\dagger \big[ {-i Z}, \Upsilon_R \big] R \big\rangle \\
    &\quad + \big\langle \mu_1^{(2)}, R^\dagger \big[ {-i \left( u_1 X + u_2 Y \right)}, \Upsilon_R \big] R \big\rangle.
\end{align*}
The symplectic form $\omega$ at the same point $(x,\mu)$ evaluated on the tangent vectors
\begin{align*}
    \xi = \Big( \Xi_R, \Xi_1^{(1)}, \Xi_1^{(2)}, \alpha_R, \alpha_1^{(1)}, \alpha_1^{(2)} \Big) & \in T_{(x,\mu)}(T^* \cM) \\
    \eta = \Big( \Upsilon_R, \Upsilon_1^{(1)}, \Upsilon_1^{(2)}, \beta_R, \beta_1^{(1)},\beta_1^{(2)} \Big) & \in T_{(x,\mu)}(T^* \cM)
\end{align*}
is given by
\begin{align*}
    \omega_{(x,\mu)}(\xi,\eta) &= \big\langle \beta_R, \Xi_R \big\rangle + \big\langle \beta_1^{(1)}, \Xi_1^{(1)} \big\rangle + \big\langle \beta_1^{(1)}, \Xi_1^{(2)} \big\rangle \\
    &\quad - \big\langle \alpha_R, \Upsilon_R \big\rangle - \big\langle \alpha_1^{(1)}, \Upsilon_1^{(1)} \big\rangle - \big\langle \alpha_1^{(2)}, \Upsilon_1^{(2)} \big\rangle \\
    &\quad - \big\langle \mu_R, [\Xi_R, \Upsilon_R] \big\rangle.
\end{align*}
Since $\dif \cH_{(x,\mu)}(\eta) = \omega_{(x,\mu)}(\xi, \eta)$ for all $\eta \in T_{(x,\mu)}(T^* \cM)$, we can match terms to get the state and costate dynamics
\begin{alignat*}{2}
    \dot{R} R^\dagger &= \Xi_R &&= {-i \left( u_1 X + u_2 Y \right)} \\
    \dot{\Omega}_1^{(1)} &= \Xi_1^{(1)} &&= {-i R^\dagger Z R} \\
    \dot{\Omega}_1^{(2)} &= \Xi_1^{(2)} &&= {-i R^\dagger \left( u_1 X + u_2 Y \right) R} \\
    \dot{\mu}_R &= \alpha_R &&= \big[ {-i \left( u_1 X + u_2 Y \right)}, \mu_R \big] \\
    & &&\quad + \big[ {-i Z}, R \mu_1^{(1)} R^\dagger \big] \\
    & &&\quad + \big[ {-i \left( u_1 X + u_2 Y \right)}, R \mu_1^{(2)} R^\dagger \big] \\
    \dot{\mu}_1^{(1)} &= \alpha_1^{(1)} &&= 0 \\
    \dot{\mu}_1^{(2)} &= \alpha_1^{(2)} &&= 0.
\end{alignat*}
The optimal controls satisfy
\begin{alignat*}{3}
    u_1^* &= \argmax_{u_1} \cH &&= \big\langle \mu_R, {-i X} \big\rangle &&+ \big\langle \mu_1^{(2)}, {-i R^\dagger X R} \big\rangle \\
    u_2^* &= \argmax_{u_2} \cH &&= \big\langle \mu_R, {-i Y} \big\rangle &&+ \big\langle \mu_1^{(2)}, {-i R^\dagger Y R} \big\rangle.
\end{alignat*}
Now we seek the initial costate
\begin{equation*}
    \mu_R(0),\ \mu_1^{(1)}(0),\ \mu_1^{(2)}(0) \in \su(2)    
\end{equation*}
such that the final state satisfies the constraints
\begin{equation*}
    R(T) = G,\ \Omega_1^{(1)}(T) = 0,\ \Omega_1^{(2)}(T) = 0    
\end{equation*}
and $u_1^*$ and $u_2^*$ are global minimizers of $J$. To remove discontinuities in the control envelopes at the initial and final times, we can apply an integrator to each control, described by the dynamics
\begin{equation*}
    \dot{u}_1 = v_1,\quad\quad \dot{u}_2 = v_2
\end{equation*}
with zero initial and final constraints ${u_1(0) = 0}$, ${u_2(0) = 0}$, ${u_1(T) = 0}$, ${u_2(T) = 0}$. We need to modify the cost functional to reflect the new controls $v_1$ and $v_2$, so let
\begin{equation*}
    J = \frac{1}{2} \int_0^T R_u \left( u_1(t)^2 + u_2(t)^2 \right) + R_v \left( v_1(t)^2 + v_2(t)^2 \right) \dif t.
\end{equation*}
where $R_u$ penalizes the squared magnitude of the control envelopes and $R_v$ penalizes the squared magnitude of the first derivatives of the control envelopes. The costate equations for the integrators are given by
\begin{alignat*}{3}
    \dot{\mu}_{u_1} &= -\langle \mu_R, {-i X} \big\rangle &&- \big\langle \mu_1^{(2)}, {-i R^\dagger X R} \big\rangle &&+ R_u u_1 \\
    \dot{\mu}_{u_2} &= -\langle \mu_R, {-i Y} \big\rangle &&- \big\langle \mu_1^{(2)}, {-i R^\dagger Y R} \big\rangle &&+ R_u u_2,
\end{alignat*}
and the new optimal controls satisfy
\begin{alignat*}{2}
    v_1^* &= \argmax_{v_1} \cH &&= \frac{1}{R_v} \mu_{u_1} \\
    v_2^* &= \argmax_{v_2} \cH &&= \frac{1}{R_v} \mu_{u_2}.
\end{alignat*}

\subsection*{Example 2}

Consider another single-qubit system with a single in-phase control ($m = 1$) subject to dephasing error only ($n = 1$), so the total Hamiltonian is given by
\begin{equation*}
    H(t) = u(t) X + \epsilon Z.
\end{equation*}
Here we will demonstrate how to suppress the disturbance to higher order ($r = 3$). We seek $u$ to implement a $\sqrt{X}$ gate
\begin{equation*}
    G = \frac{1}{\sqrt{2}} \begin{bmatrix} 1 & -i \\ -i & 1 \end{bmatrix}
\end{equation*}
while minimizing the cost functional
\begin{equation*}
    J = \frac{1}{2} \int_0^T u(t)^2 \dif t.
\end{equation*}
The control Hamiltonian is given by
\begin{align*}
    \cH &= \big\langle \mu_R, {-i u X} \big\rangle + \big\langle \mu_1^{(1)}, {-i R^\dagger Z R} \big\rangle \\
    &\quad + \big\langle \mu_2^{(1,1)}, {-\frac{1}{2}} \big[ \Omega_1^{(1)}, {-i R^\dagger Z R} \big] \big\rangle \\
    &\quad + \big\langle \mu_3^{(1,1,1)}, {-\frac{1}{2}} \big[ \Omega_2^{(1,1)}, {-i R^\dagger Z R} \big] \\
    &\quad + \frac{1}{12} \big[ \Omega_1^{(1)}, \big[ \Omega_1^{(1)}, {-i R^\dagger Z R} \big] \big] \big\rangle - \frac{1}{2} u^2.
\end{align*}
Evaluating the differential of $\cH$ and the symplectic form $\omega$ at the point
\begin{alignat*}{2}
    (x,\mu) &= \Big(R, \Omega_1^{(1)}, \Omega_2^{(1,1)}, \Omega_3^{(1,1,1)}, && \\
    &\hspace{2cm} \mu_R, \mu_1^{(1)}, \mu_2^{(1,1)}, \mu_3^{(1,1,1)} \Big) && \in T^* \cM
\end{alignat*}
on the tangent vectors
\begin{alignat*}{2}
    \xi &= \Big( \Xi_R, \Xi_1^{(1)}, \Xi_2^{(1,1)}, \Xi_3^{(1,1,1)}, && \\
    &\hspace{2cm} \alpha_R, \alpha_1^{(1)}, \alpha_2^{(1,1)}, \alpha_3^{(1,1,1)} \Big) && \in T_{(x,\mu)}(T^* \cM) \\
    \eta &= \Big( \Upsilon_R, \Upsilon^{(1)}, \Upsilon_2^{(1,1)}, \Upsilon_3^{(1,1,1)}, && \\
    &\hspace{2cm} \beta_R, \beta_1^{(1)}, \beta_2^{(1,1)}, \beta_3^{(1,1,1)} \Big) && \in T_{(x,\mu)}(T^* \cM)
\end{alignat*}
gives
{\small
\begin{align*}
    &\dif \cH_{(x,\mu)}(\eta) = \big\langle \beta_R, {-i u X} \big\rangle \\
    & + \big\langle \beta_1^{(1)}, {-i R^\dagger Z R} \big\rangle + \big\langle \beta_2^{(1,1)}, {-\frac{1}{2}} \big[ \Omega_1^{(1)}, {-i R^\dagger Z R} \big] \big\rangle \\
    & + \big\langle \beta_3^{(1,1,1)}, {-\frac{1}{2}} \big[ \Omega_2^{(1,1)}, {-i R^\dagger Z R} \big] + \frac{1}{12} \big[ \Omega_1^{(1)}, \big[ \Omega_1^{(1)}, {-i R^\dagger Z R} \big] \big] \big\rangle \\
    & + \big\langle \mu_1^{(1)}, R^\dagger \big[ {-i Z}, \Upsilon_R \big] R \big\rangle \\
    & + \big\langle \mu_2^{(1,1)}, {-\frac{1}{2}} \big[ \Omega_1^{(1)}, R^\dagger \big[ {-i Z}, \Upsilon_R \big] R \big] - \frac{1}{2} \big[ \Upsilon_1^{(1)}, {-i R^\dagger Z R} \big] \big\rangle \\
    & + \big\langle \mu_3^{(1,1,1)}, {-\frac{1}{2}} \big[ \Omega_2^{(1,1)}, R^\dagger \big[ {-i Z}, \Upsilon_R \big] R \big] \\
    & + \frac{1}{12} \big[ \Omega_1^{(1)}, \big[ \Omega_1^{(1)}, R^\dagger \big[ {-i Z}, \Upsilon_R \big] R \big] \big] - \frac{1}{2} \big[ \Upsilon_2^{(1,1)}, {-i R^\dagger Z R} \big] \\
    & + \frac{1}{12} \big[ \Upsilon_1^{(1)}, \big[ \Omega_1^{(1)}, {-i R^\dagger Z R} \big] \big] + \frac{1}{12} \big[ \Omega_1^{(1)}, \big[ \Upsilon_1^{(1)}, {-i R^\dagger Z R} \big] \big] \big\rangle
\end{align*}}
and
\begin{align*}
    &\omega_{(x,\mu)}(\xi,\eta) = \big\langle \beta_R, \Xi_R \big\rangle + \big\langle \beta_1^{(1)}, \Xi_1^{(1)} \big\rangle + \big\langle \beta_2^{(1,1)}, \Xi_2^{(1,1)} \big\rangle \\
    & + \big\langle \beta_3^{(1,1,1)}, \Xi_3^{(1,1,1)} \big\rangle - \big\langle \alpha_R, \Upsilon_R \big\rangle - \big\langle \alpha_1^{(1)}, \Upsilon_1^{(1)} \big\rangle \\
    & - \big\langle \alpha_2^{(1,1)}, \Upsilon_2^{(1,1)} \big\rangle - \big\langle \alpha_3^{(1,1,1)}, \Upsilon_3^{(1,1,1)} \big\rangle - \big\langle \mu_R, [\Xi_R, \Upsilon_R] \big\rangle.
\end{align*}
Matching terms yields the dynamics
{\small
\begin{align*}
    \dot{R} R^\dagger &= {-i u X} \\
    \dot{\Omega}_1^{(1)} &= {-i R^\dagger Z R} \\
    \dot{\Omega}_2^{(1,1)} &= {-\frac{1}{2}} \big[ \Omega_1^{(1)}, {-i R^\dagger Z R} \big] \\
    \dot{\Omega}_3^{(1,1,1)} &= {-\frac{1}{2}} \big[ \Omega_2^{(1,1)}, {-i R^\dagger Z R} \big] + \frac{1}{12} \big[ \Omega_1^{(1)}, \big[ \Omega_1^{(1)}, {-i R^\dagger Z R} \big] \big] \\
    \dot{\mu}_R &= \big[ {-i u X}, \mu_R \big] + \big[ {-i Z}, R \Big( \mu_1^{(1)} - \frac{1}{2} \big[ \mu_2^{(1,1)}, \Omega_1^{(1)} \big] \\
    &\quad - \frac{1}{2} \big[ \mu_3^{(1,1,1)}, \Omega_2^{(1,1)} \big] + \frac{1}{12} \big[ \big[ \mu_3^{(1,1,1)}, \Omega_1^{(1)} \big], \Omega_1^{(1)} \big] \Big) R^\dagger \big] \\
    \dot{\mu}_1^{(1)} &= \frac{1}{2} \big[ {-i R^\dagger Z R}, \mu_2^{(1,1)} \big] - \frac{1}{12} \big[ \big[ \Omega_1^{(1)}, {-i R^\dagger Z R} \big], \mu_3^{(1,1,1)} \big] \\
    &\quad - \frac{1}{12} \big[ {-i R^\dagger Z R}, \big[ \Omega_1^{(1)}, \mu_3^{(1,1,1)} \big] \big] \\
    \dot{\mu}_2^{(1,1)} &= \frac{1}{2} \big[ {-i R^\dagger Z R}, \mu_3^{(1,1,1)} \big] \\
    \dot{\mu}_3^{(1,1,1)} &= 0,
\end{align*}}
and the optimal control satisfies
\begin{equation*}
    u^* = \argmax_{u} \cH = \big\langle \mu_R, {-i X} \big\rangle.
\end{equation*}
As in the previous example, we seek the initial costate
\begin{equation*}
    \mu_R(0),\ \mu_1^{(1)}(0),\ \mu_2^{(1,1)}(0),\ \mu_3^{(1,1,1)}(0) \in \su(2)    
\end{equation*}
such that
\begin{equation*}
    R(T) = G,\ \Omega_1^{(1)}(T) = 0,\ \Omega_2^{(1,1)}(T) = 0,\ \Omega_3^{(1,1,1)}(T) = 0
\end{equation*}
and $u^*$ globally minimizes $J$. If greater smoothness in the control envelopes is desired, we may apply cascaded integrators to the controls and introduce additional costate equations as described in the previous example.

\section{Numerical results}\label{sec:results}

In this section we solve the optimization problems stated in Examples 1 and 2 and plot the resulting control pulses, state trajectories, and gate infidelities. Using two single-qubit examples we apply the framework to one setting with two independent noise sources and to another setting with one noise source which is to be suppressed to third order. Each example applies a single integrator to the control input(s) so that we can constrain the initial and final values of the pulse envelopes to zero. The cost functional for both examples is given by the integral of the squared magnitude of the first derivative of the controls, i.e., $R_u = 0$ and $R_v = 1$.

\begin{figure}
    \centering
    \includegraphics[width=1.0\linewidth]{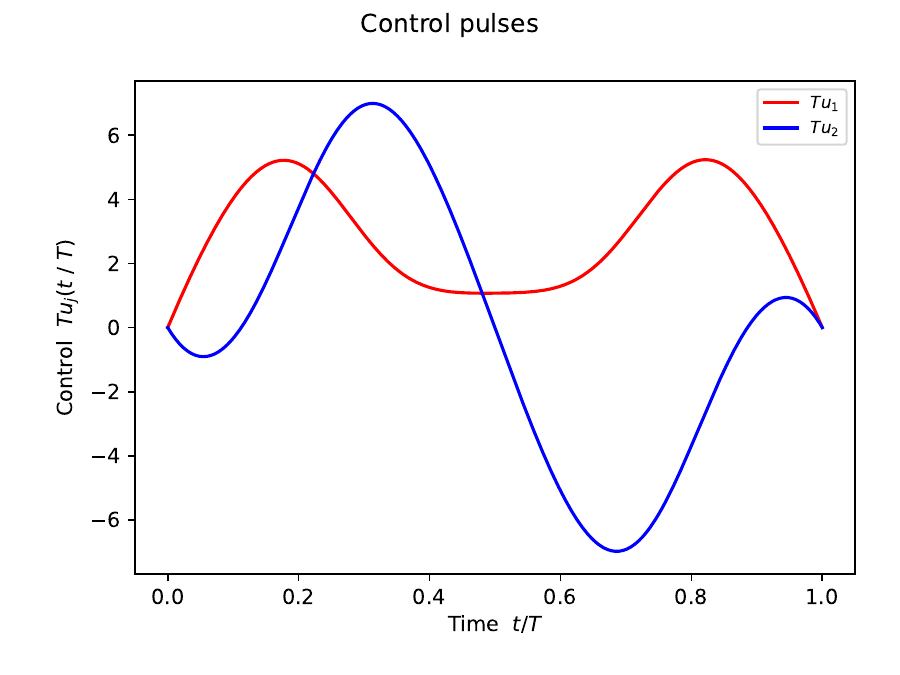}
    \caption{In-phase and quadrature components ($u_1$ and $u_2$) of a pulse envelope implementing a Hadamard gate that is first-order robust to dephasing and multiplicative control errors.}
    \label{fig:ex1_control}
\end{figure}

\begin{figure}
    \centering
    \includegraphics[width=1.0\linewidth]{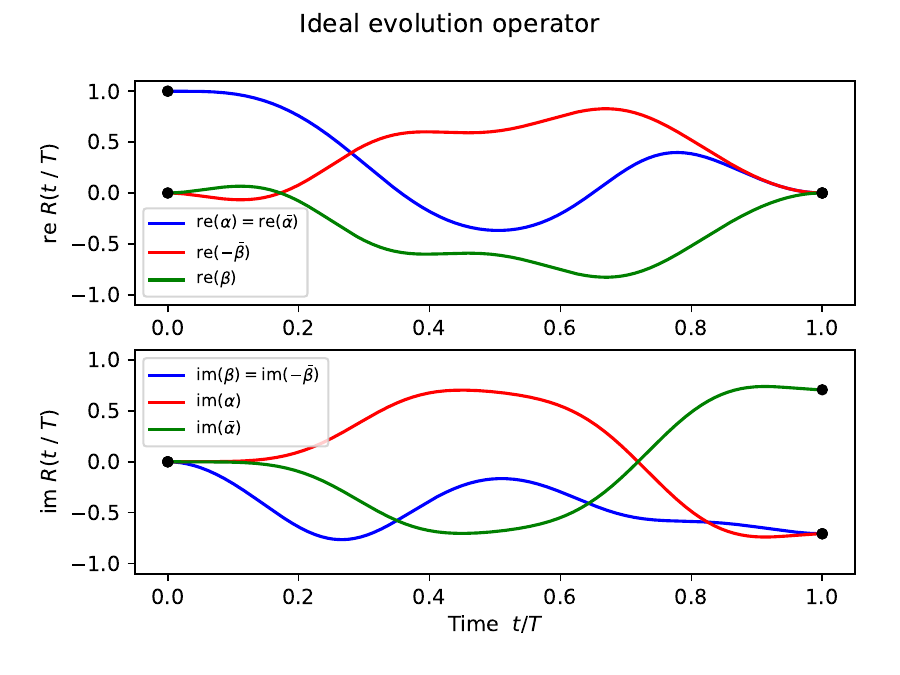}
    \caption{Trajectory of the ideal evolution operator implementing a Hadamard gate that is first-order robust to dephasing and multiplicative control errors.}
    \label{fig:ex1_evolution}
\end{figure}

\begin{figure*}
    \centering
    \begin{subfigure}[b]{0.32\textwidth}
        \centering
        \includegraphics[width=\linewidth]{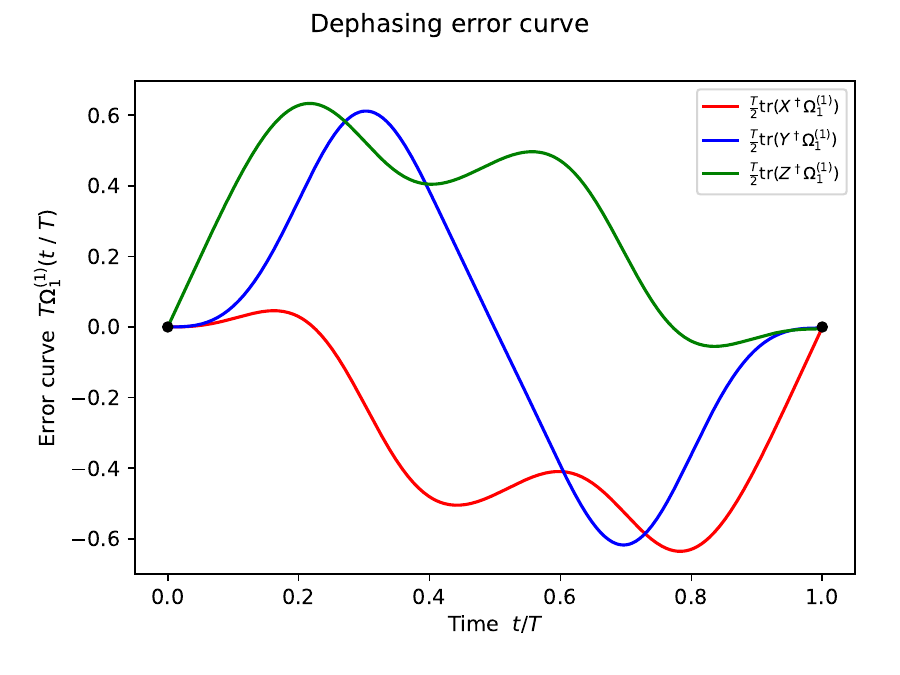}
        \caption{Trajectory of the first-order error curve for the dephasing disturbance, projected onto the standard $\su(2)$ basis.}
        \label{fig:ex1_dephasing_error}
    \end{subfigure}
    \hspace{1cm}
    \begin{subfigure}[b]{0.32\textwidth}
        \centering
        \includegraphics[width=\linewidth]{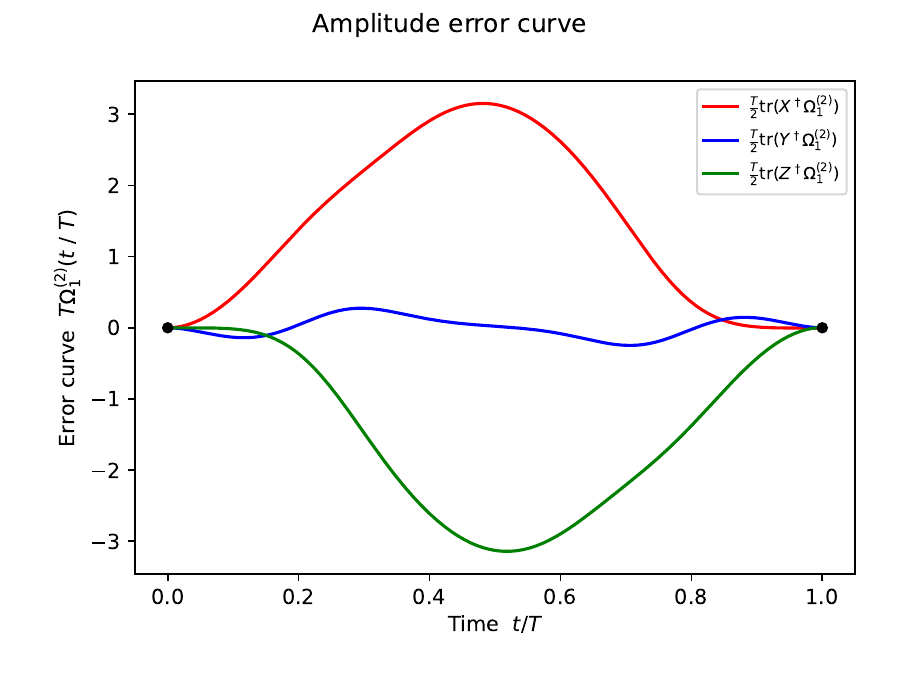}
        \caption{Trajectory of the first-order error curve for the amplitude disturbance, projected onto the standard $\su(2)$ basis.}
        \label{fig:ex1_amplitude_error}
    \end{subfigure}
    \vskip\baselineskip
    \begin{subfigure}[b]{0.32\textwidth}
        \centering
        \includegraphics[trim={2cm 1cm 2cm 2cm},clip,width=\linewidth]{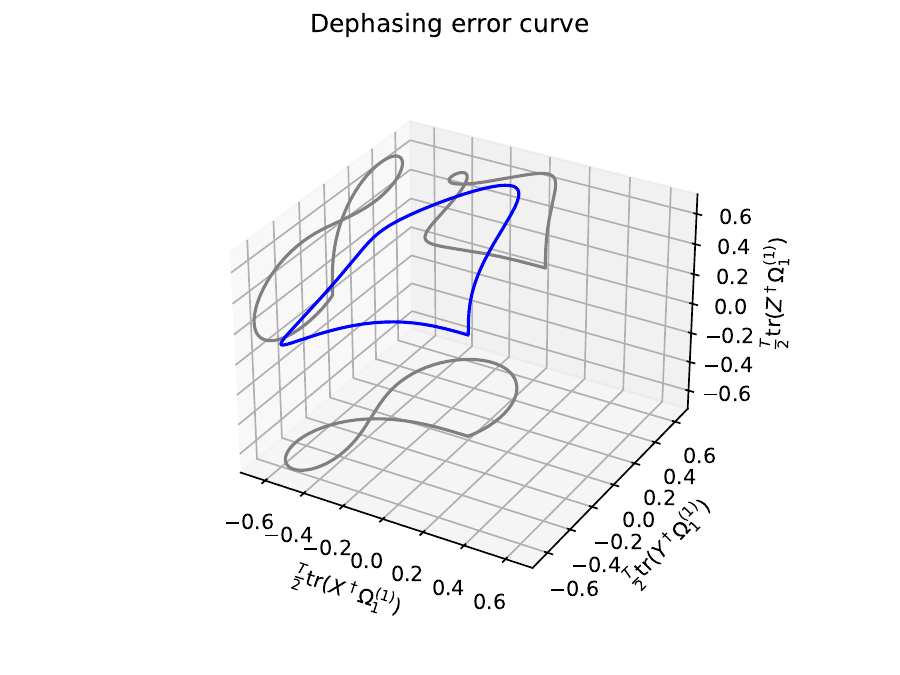}
        \caption{Closed first-order error curve for the dephasing disturbance.}
        \label{fig:ex1_dephasing_error_curve}
    \end{subfigure}
    \hspace{1cm}
    \begin{subfigure}[b]{0.32\textwidth}
        \centering
        \includegraphics[trim={2cm 1cm 2cm 2cm},clip,width=\linewidth]{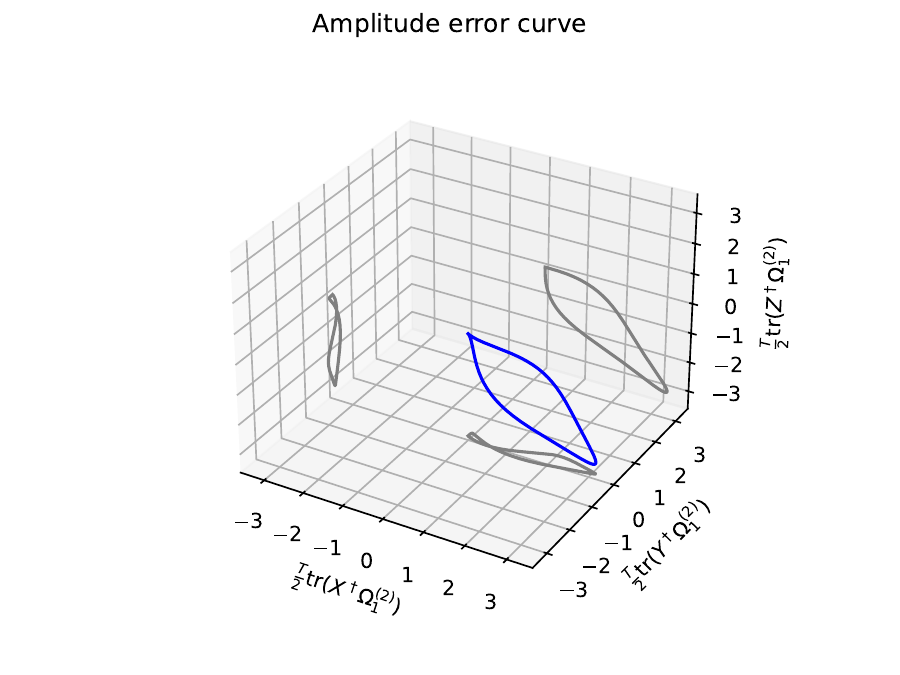}
        \caption{Closed first-order error curve for the amplitude disturbance.}
        \label{fig:ex1_amplitude_error_curve}
    \end{subfigure}
    \caption{First-order dephasing and amplitude error curves for the Hadamard gate.}
    \label{fig:ex1_error_curves}
\end{figure*}

\begin{figure*}
    \centering
    \begin{subfigure}[b]{0.32\textwidth}
        \centering
        \includegraphics[width=\linewidth]{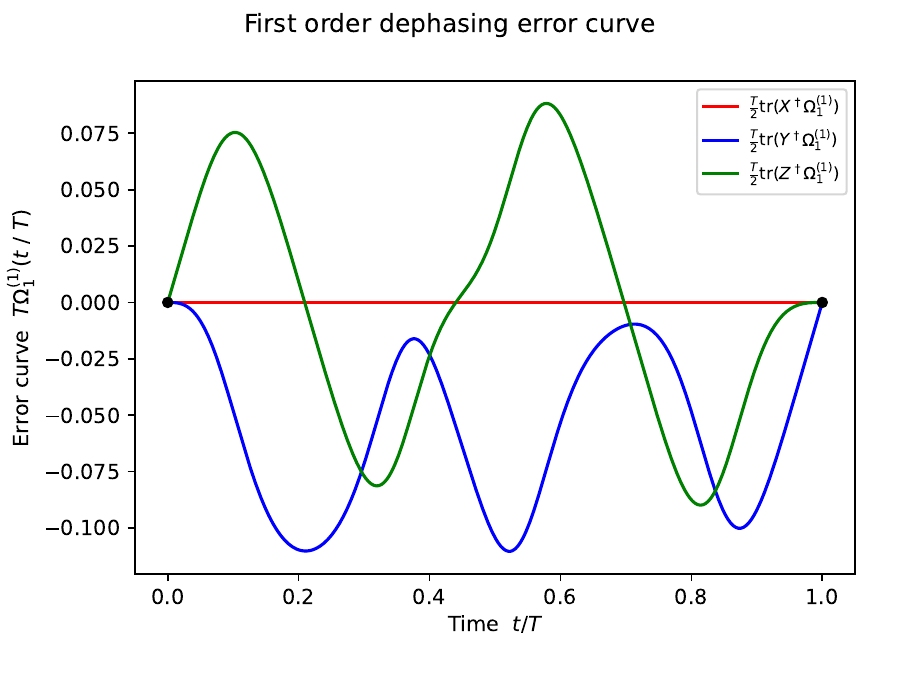}
        \caption{Trajectory of the first-order error curve for the dephasing disturbance, projected onto the standard $\su(2)$ basis.}
        \label{fig:ex2_first_order_dephasing_error}
    \end{subfigure}
    \hfill
    \begin{subfigure}[b]{0.32\textwidth}
        \centering
        \includegraphics[width=\linewidth]{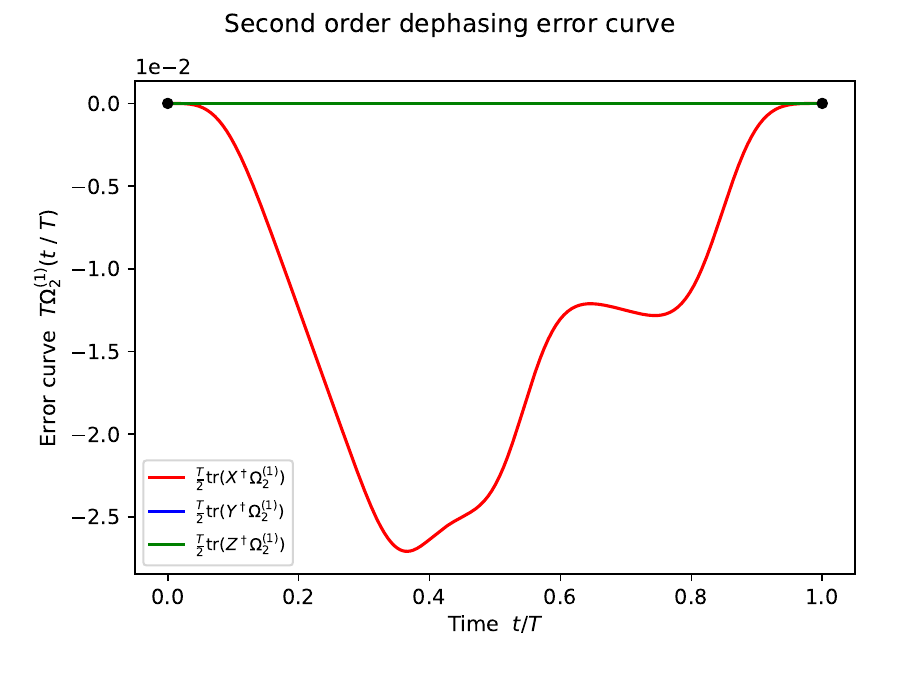}
        \caption{Trajectory of the second-order error curve for the dephasing disturbance, projected onto the standard $\su(2)$ basis.}
        \label{fig:ex2_second_order_dephasing_error}
    \end{subfigure}
    \hfill
    \begin{subfigure}[b]{0.32\textwidth}
        \centering
        \includegraphics[width=\linewidth]{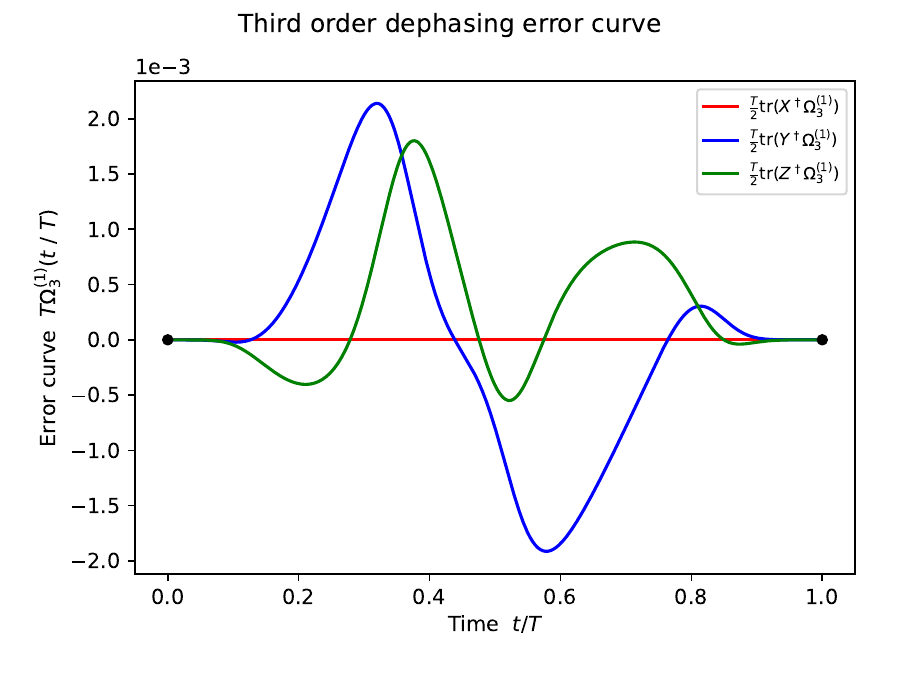}
        \caption{Trajectory of the third-order error curve for the dephasing disturbance, projected onto the standard $\su(2)$ basis.}
        \label{fig:ex2_third_order_dephasing_error}
    \end{subfigure}
    \vskip\baselineskip
    \begin{subfigure}[b]{0.32\textwidth}
        \centering
        \includegraphics[trim={2cm 1cm 2cm 2cm},clip,width=\linewidth]{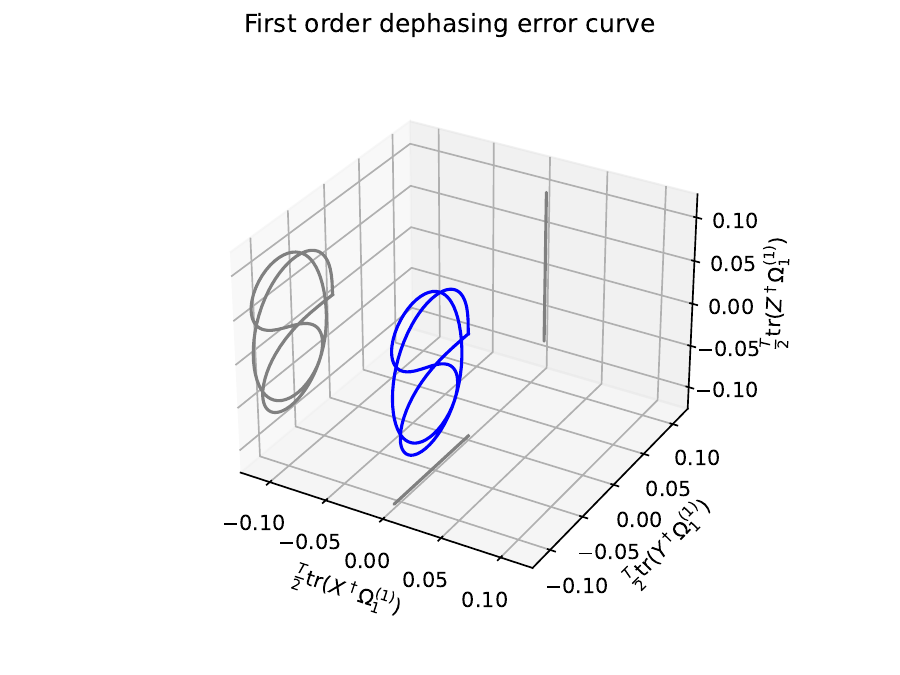}
        \caption{Closed first-order error curve for the dephasing disturbance.}
        \label{fig:ex2_first_order_dephasing_error_curve}
    \end{subfigure}
    \hfill
    \begin{subfigure}[b]{0.32\textwidth}
        \centering
        \includegraphics[trim={2cm 1cm 2cm 2cm},clip,width=\linewidth]{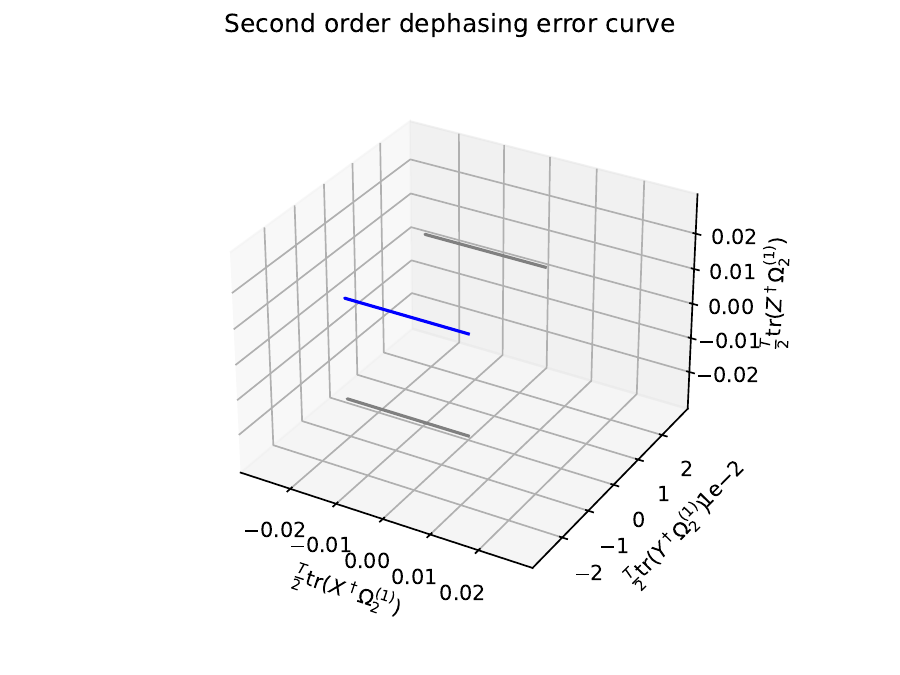}
        \caption{Closed second-order error curve for the dephasing disturbance.}
        \label{fig:ex2_second_order_dephasing_error_curve}
    \end{subfigure}
    \hfill
    \begin{subfigure}[b]{0.32\textwidth}
        \centering
        \includegraphics[trim={2cm 1cm 2cm 2cm},clip,width=\linewidth]{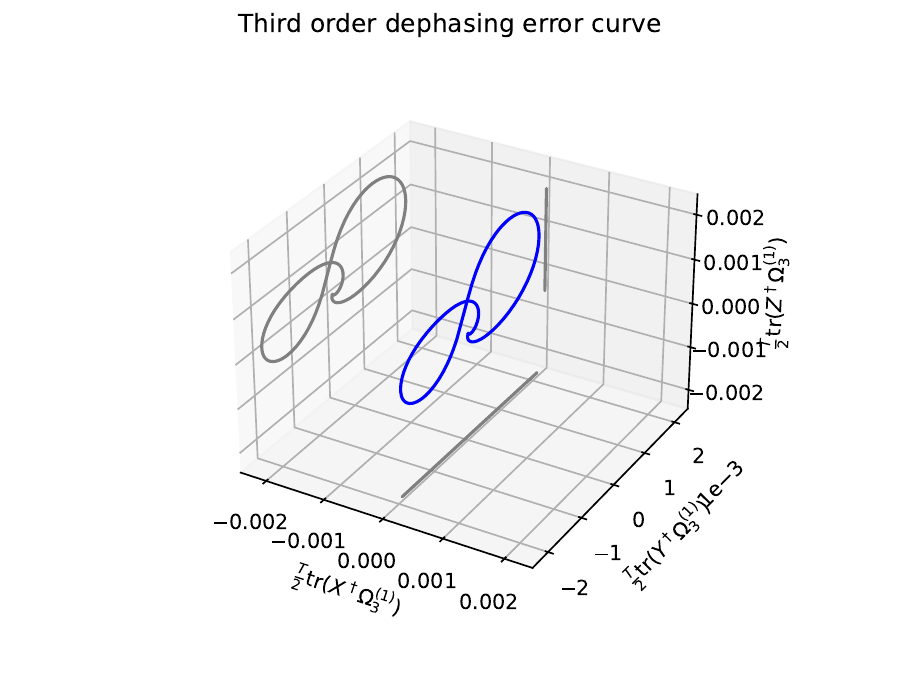}
        \caption{Closed third-order error curve for the dephasing disturbance.}
        \label{fig:ex2_third_order_dephasing_error_curve}
    \end{subfigure}
    \caption{First-, second-, and third-order dephasing error curves for the $\sqrt{X}$ gate.}
    \label{fig:ex2_error_curves}
\end{figure*}

The optimal controls for Examples 1 and 2 are plotted in Figures \ref{fig:ex1_control} and \ref{fig:ex2_control}, respectively, with the corresponding gate infidelities as a function of the disturbance magnitudes plotted in Figures \ref{fig:ex1_infidelity} and \ref{fig:ex2_infidelity}. In Figures \ref{fig:ex1_evolution} and \ref{fig:ex2_evolution} we plot the trajectory of the ideal unitary evolution operator. The final value represents the unitary gate implemented by the control pulse in the limit of vanishing disturbances. We parameterize the time-varying evolution operator $R : [0,T] \to \SU(2)$ as
\begin{equation*}
    R(t) = \begin{bmatrix} \alpha(t) & -\bar{\beta}(t) \\ \beta(t) & \bar{\alpha}(t) \end{bmatrix}
\end{equation*}
for complex numbers $\alpha,\beta : [0,T] \to \Complex$ satisfying the constraint $|\alpha(t)|^2 + |\beta(t)|^2 = 1$ at each time $t \in [0,T]$, where $\bar{\cdot}$ denotes the complex conjugate. The error curves governed by \eqref{eq:Omega_dynamics} are plotted in Figures \ref{fig:ex1_error_curves} and \ref{fig:ex2_error_curves}. We illustrate both the time-dependent trajectories and the three-dimensional closed space curves in the Lie algebra.

\begin{figure}
    \centering
    \includegraphics[width=1.0\linewidth]{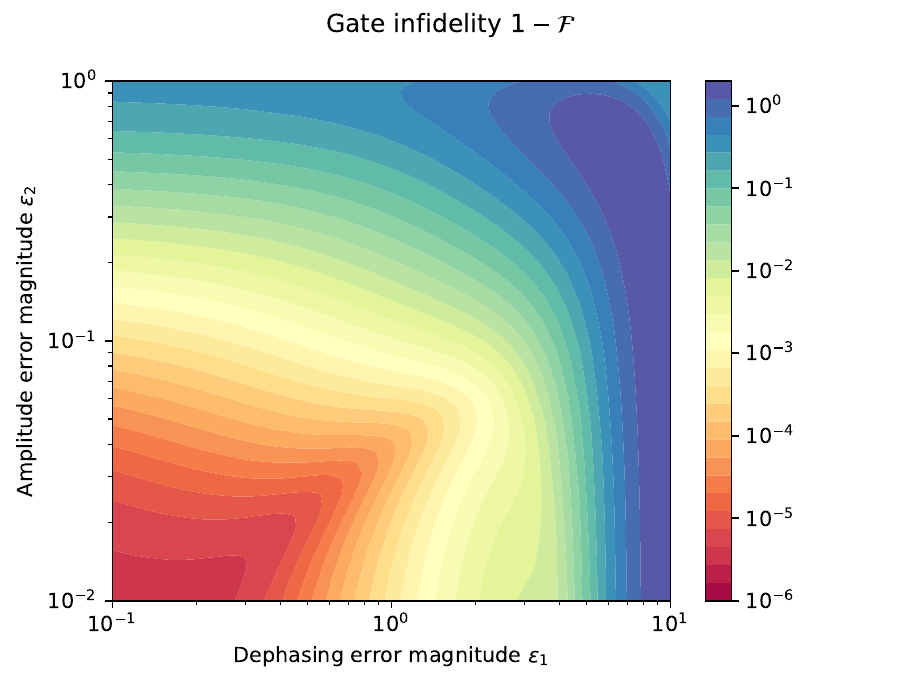}
    \caption{Gate infidelity as a function of the two disturbance magnitudes $\epsilon_1$ and $\epsilon_2$ for the controls shown in Figure \ref{fig:ex1_control}. The vanishing noise limit is in the lower-left region.}
    \label{fig:ex1_infidelity}
\end{figure}

All of the differential equations are simulated using the fourth-order Runge-Kutta method with absolute and relative tolerances set to $10^{-8}$. We solve the transcribed optimization algorithm using a trust region algorithm, with gradients of the objective function computed using automatic differentiation functionality in Julia. Notice that the infidelity in Figure \ref{fig:ex2_infidelity} does not exactly go to zero as the disturbance magnitudes vanish due to limited numerical precision related to the discretization algorithm used in the ODE solvers and floating point round-off error. This can be mitigated by using geometric numerical integrators designed specifically for ODEs on Lie groups (e.g., Runge-Kutta-Munthe-Kaas methods \cite{munthekaas1998rungekutta}), which affects both the accuracy of the ODE solutions and the gradients of the objective function in the transcribed optimization problem.

\begin{figure}
    \centering
    \includegraphics[width=1.0\linewidth]{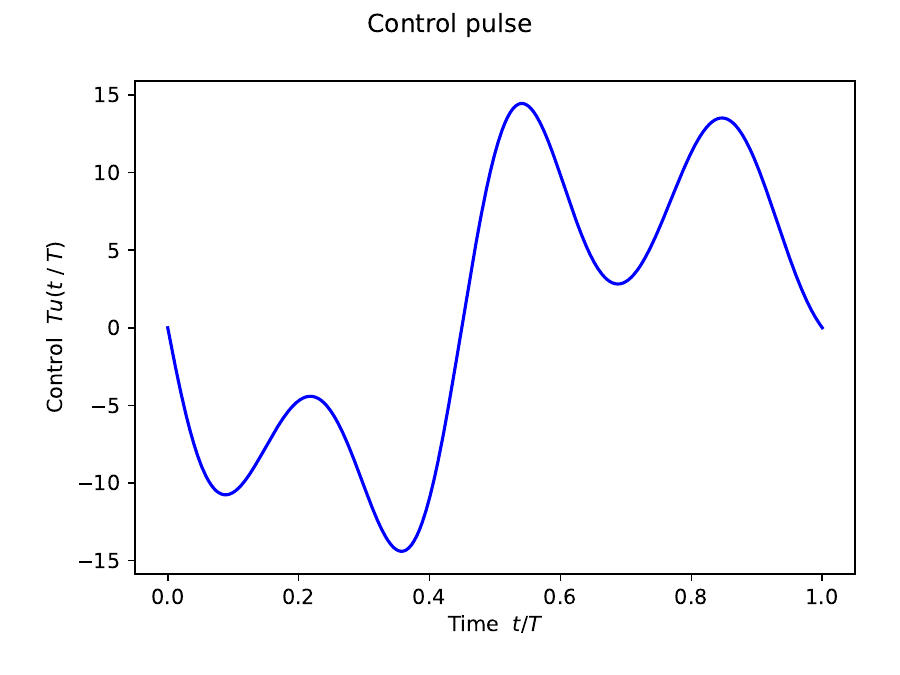}
    \caption{In-phase pulse envelope implementing a $\sqrt{X}$ gate that is third-order robust to dephasing.}
    \label{fig:ex2_control}
    \setcounter{figure}{5}
\end{figure}

\begin{figure}
    \centering
    \includegraphics[width=1.0\linewidth]{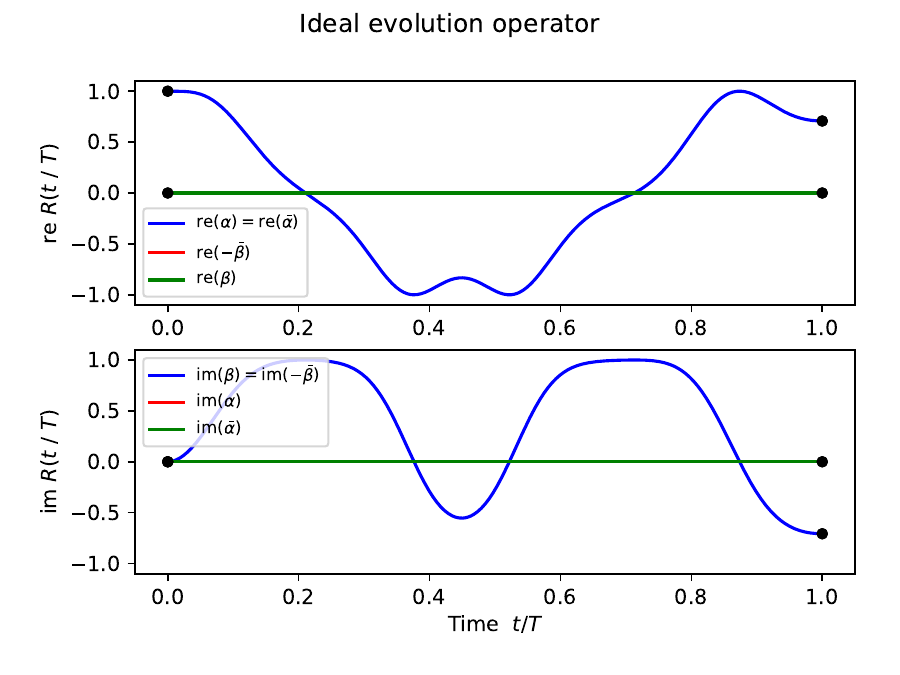}
    \caption{Trajectory of the ideal evolution operator implementing a $\sqrt{X}$ gate that is third-order robust to dephasing.}
    \label{fig:ex2_evolution}
    \setcounter{figure}{6}
\end{figure}

\begin{figure}
    \centering
    \includegraphics[width=1.0\linewidth]{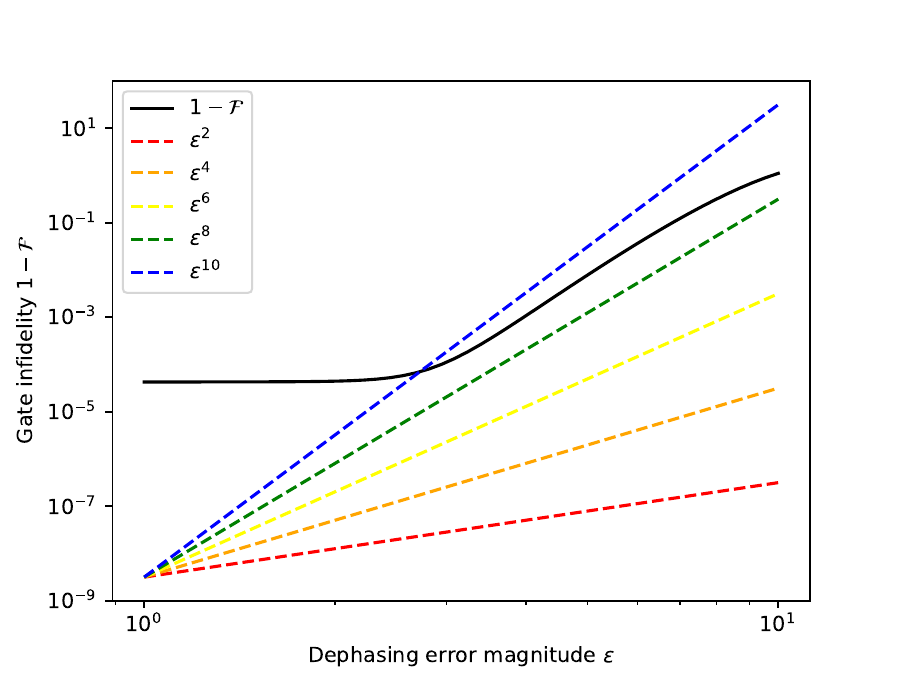}
    \caption{Gate infidelity as a function of the disturbance magnitude $\epsilon$ for the control shown in Figure \ref{fig:ex2_control}.}
    \label{fig:ex2_infidelity}
\end{figure}

\section{Discussion}\label{sec:discussion}

In this work we have formulated a general purpose framework for synthesizing smooth noise-robust controls for unitary quantum dynamics based on Pontryagin's maximum principle. The approach is applicable to systems with any number of controls and disturbances, and any number of energy levels, subject to reasonable reachability constraints. Using two single-qubit examples, we apply the framework to a setting with two independent noise sources to be suppressed to first-order and to a setting with one noise source to be suppressed to third-order. Each example applies a single integrator to the control input(s) so that we can constrain the initial and final values of the pulse envelopes to zero. Higher order derivative constraints on the controls can be enforced by applying additional integrators with corresponding initial and final state constraints. In the same manner that artificial integrators were included to impose smoothness constraints, dynamical distortion models for the control electronics can also be compensated for intrinsically by including them in the augmented control system.

The maximum principle allows us to encode entire pulse envelopes using only the initial costate, which can be efficiently represented as a finite-dimensional vector of floating point numbers. We can reconstruct the control to any desired precision by solving Hamilton's equations from PMP and sampling the solution trajectory. The set of initial costates for a parameterized family of noise-suppressing gates forms a submanifold of $\su(N)^{p+1}$. Learning a smooth parameterization of this manifold via e.g., splines or neural nets from a finite point cloud enables \textit{post hoc} interpolation between noise-robust gates within this parameterized family, which can be useful for performing e.g., non-standard rotations or \textit{in situ} gate recalibration.

Leakage to energy levels outside the computational basis is another important phenomenon which often needs to be mitigated, especially in manufactured qubits such as transmons. We can design leakage-suppressing pulse envelopes using this framework by replacing the final endpoint constraint in the optimal control problem with an appropriate final set constraining only the submatrices of the evolution operator and error curves corresponding to the computational subsystem as well as the off-diagonal blocks, leaving the submatrices for the non-computational subspace as degrees of freedom. The transversality condtions in PMP requires the optimal control policy to drive the costate to a final value that is orthogonal to this final set, replacing some number of final state constraints with an equal number of final costate constraints. The optimization algorithms for finding the initial costate then are very similar except for a slight modification to the objective function to reflect the additional degrees of freedom on the final state and the additional constraints on the final costate. Developing optimal leakage-suppressing pulses and comparing them to pulses obtained using the Derivative Removal by Adiabatic Gate (DRAG) \cite{motzoi2009simple} strategy and its higher order extensions is a direction for future work.

\section*{Acknowledgment}

This work was supported by the U.S. Department of Energy, Office of SBIR/STTR Programs under Award Number DE-SC0022389.

\bibliographystyle{IEEEtran}
\bibliography{references}

\begin{thebibliography}{10}
\providecommand{\url}[1]{#1}
\csname url@samestyle\endcsname
\providecommand{\newblock}{\relax}
\providecommand{\bibinfo}[2]{#2}
\providecommand{\BIBentrySTDinterwordspacing}{\spaceskip=0pt\relax}
\providecommand{\BIBentryALTinterwordstretchfactor}{4}
\providecommand{\BIBentryALTinterwordspacing}{\spaceskip=\fontdimen2\font plus
\BIBentryALTinterwordstretchfactor\fontdimen3\font minus
  \fontdimen4\font\relax}
\providecommand{\BIBforeignlanguage}[2]{{%
\expandafter\ifx\csname l@#1\endcsname\relax
\typeout{** WARNING: IEEEtran.bst: No hyphenation pattern has been}%
\typeout{** loaded for the language `#1'. Using the pattern for}%
\typeout{** the default language instead.}%
\else
\language=\csname l@#1\endcsname
\fi
#2}}
\providecommand{\BIBdecl}{\relax}
\BIBdecl

\bibitem{viola1999dynamical}
L.~Viola, E.~Knill, and S.~Lloyd, ``Dynamical decoupling of open quantum
  systems,'' \emph{Physical Review Letters}, vol.~82, no.~12, p. 2417, 1999.

\bibitem{khodjasteh2009dynamically}
K.~Khodjasteh and L.~Viola, ``Dynamically error-corrected gates for universal
  quantum computation,'' \emph{Physical review letters}, vol. 102, no.~8, p.
  080501, 2009.

\bibitem{zeng2019geometric}
J.~Zeng, C.~Yang, A.~Dzurak, and E.~Barnes, ``Geometric formalism for
  constructing arbitrary single-qubit dynamically corrected gates,''
  \emph{Physical Review A}, vol.~99, no.~5, p. 052321, 2019.

\bibitem{buterakos2021geometrical}
D.~Buterakos, S.~D. Sarma, and E.~Barnes, ``Geometrical formalism for
  dynamically corrected gates in multiqubit systems,'' \emph{PRX Quantum},
  vol.~2, no.~1, p. 010341, 2021.

\bibitem{khaneja2005optimal}
N.~Khaneja, T.~Reiss, C.~Kehlet, T.~Schulte-Herbr{\"u}ggen, and S.~J. Glaser,
  ``Optimal control of coupled spin dynamics: design of nmr pulse sequences by
  gradient ascent algorithms,'' \emph{Journal of magnetic resonance}, vol. 172,
  no.~2, pp. 296--305, 2005.

\bibitem{pupillo2022timeoptimal}
\BIBentryALTinterwordspacing
S.~Jandura and G.~Pupillo, ``Time-{O}ptimal {T}wo- and {T}hree-{Q}ubit {G}ates
  for {R}ydberg {A}toms,'' \emph{{Quantum}}, vol.~6, p. 712, May 2022.
  [Online]. Available: \url{https://doi.org/10.22331/q-2022-05-13-712}
\BIBentrySTDinterwordspacing

\bibitem{barnes2022dynamically}
E.~Barnes, F.~A. Calderon-Vargas, W.~Dong, B.~Li, J.~Zeng, and F.~Zhuang,
  ``Dynamically corrected gates from geometric space curves,'' \emph{Quantum
  Science and Technology}, vol.~7, no.~2, p. 023001, 2022.

\bibitem{magnus1954exponential}
W.~Magnus, ``{On the exponential solution of differential equations for a
  linear operator},'' \emph{Commun. Pure Appl. Math.}, vol.~7, pp. 649--673,
  1954.

\bibitem{SKINNER2010248}
T.~E. Skinner and N.~I. Gershenzon, ``Optimal control design of pulse shapes as
  analytic functions,'' \emph{Journal of Magnetic Resonance}, vol. 204, no.~2,
  pp. 248--255, 2010.

\bibitem{lucarelli2018quantum}
D.~Lucarelli, ``Quantum optimal control via gradient ascent in function space
  and the time-bandwidth quantum speed limit,'' \emph{Physical Review A},
  vol.~97, no.~6, p. 062346, 2018.

\bibitem{oda2023optimally}
Y.~Oda, D.~Lucarelli, K.~Schultz, B.~D. Clader, and G.~Quiroz, ``Optimally
  band-limited noise filtering for single-qubit gates,'' \emph{Physical Review
  Applied}, vol.~19, no.~1, p. 014062, 2023.

\bibitem{sorensen2018quantum}
J.~J.~W. S{\o}rensen, M.~Aranburu, T.~Heinzel, and J.~F. Sherson, ``Quantum
  optimal control in a chopped basis: Applications in control of bose-einstein
  condensates,'' \emph{Physical Review A}, vol.~98, no.~2, p. 022119, 2018.

\bibitem{machnes2018tunable}
S.~Machnes, E.~Ass{\'e}mat, D.~Tannor, and F.~K. Wilhelm, ``Tunable, flexible,
  and efficient optimization of control pulses for practical qubits,''
  \emph{Physical review letters}, vol. 120, no.~15, p. 150401, 2018.

\bibitem{chang2011simple}
\BIBentryALTinterwordspacing
D.~E. Chang, ``A simple proof of the pontryagin maximum principle on
  manifolds,'' \emph{Automatica}, vol.~47, no.~3, pp. 630--633, 2011. [Online].
  Available:
  \url{https://www.sciencedirect.com/science/article/pii/S0005109811000525}
\BIBentrySTDinterwordspacing

\bibitem{munthekaas1998rungekutta}
H.~Munthe-Kaas, ``Runge-kutta methods on lie groups,'' \emph{BIT Numerical
  Mathematics}, vol.~38, no.~1, p. 92–111, Mar 1998.

\bibitem{motzoi2009simple}
F.~Motzoi, J.~M. Gambetta, P.~Rebentrost, and F.~K. Wilhelm, ``Simple pulses
  for elimination of leakage in weakly nonlinear qubits,'' \emph{Physical
  review letters}, vol. 103, no.~11, p. 110501, 2009.

\end{thebibliography}

\end{document}